\documentclass[letterpaper, 10 pt, conference]{ieeeconf}

\IEEEoverridecommandlockouts
\usepackage{graphics} 
\usepackage{amsmath}
\usepackage{amssymb}  
\usepackage{amsfonts} 
\usepackage{graphicx,float,wrapfig,subfigure}
\usepackage{cite}
\usepackage{bm}
\usepackage{soul,color}
\usepackage[linesnumbered,ruled]{algorithm2e}
\usepackage{url}

\title{\LARGE \bf
Real-time Ecological Velocity Planning for Plug-in Hybrid Vehicles\\ with Partial Communication to Traffic Lights}

\author{Sangjae Bae$^{1}$, Yongkeun Choi$^{2}$, Yeojun Kim$^{2}$, Jacopo Guanetti$^{2}$, Francesco Borrelli$^{2}$, and Scott Moura$^{1}$
\thanks{$^{1}$Department of Civil and Environmental Engineering, University of California, Berkeley, USA. Corresponding author: {\tt\small sangjae.bae@berkeley.edu}}%
\thanks{$^{2}$Department of Mechanical Engineering, University of California, Berkeley, USA.}}%

\begin{document}
\maketitle
\thispagestyle{empty}
\pagestyle{empty}

\begin{abstract}
This paper presents the design of an ecological adaptive cruise controller (ECO-ACC) for a plug-in hybrid vehicle (PHEV) which exploits automated driving and connectivity. Most existing papers for ECO-ACC focus on a short-sighted control scheme. A two-level control framework for long-sighted ECO-ACC was only recently introduced \cite{bae2018design}. However, that work is based on a deterministic traffic signal phase and timing (SPaT) over the entire route. In practice, connectivity with traffic lights may be limited by communication range, e.g. just one upcoming traffic light. We propose a two-level receding-horizon control framework for long-sighted ECO-ACC that exploits deterministic SPaT for the upcoming traffic light, and utilizes historical SPaT for other traffic lights within a receding control horizon. We also incorporate a powertrain control mechanism to enhance PHEV energy prediction accuracy. Hardware-in-the-loop simulation results validate the energy savings of the receding-horizon control framework in various traffic scenarios.
\end{abstract}

\section{Introduction}

Advanced Driving Assistance Systems (ADAS) are driving automation technologies that seek to improve driver comport and safety. Adaptive Cruise Control (ACC), autonomous emergency braking, and lane keeping assistance are examples of widely deployed functions in today's vehicles with so-called Level 2 automation \cite{SAE-J3016_201401}. Vehicle-to-Infrastructure (V2I) and Vehicle-to-Vehicle (V2V) connectivity further advance innovative ADAS technologies, particularly including energy-efficient driving \cite{Guanetti2018}.


To reduce energy consumption, many studies have focused on longitudinal control and proposed ecological ACC designs within short, immediate surroundings \cite{Schmied2015,turri2017model}. In that regard, given a fixed route, finding an optimal velocity trajectory, or ``Eco-driving", has been studied from the perspective of optimal control \cite{Sciarretta2015, Ozatay2014, Guanetti2018}. In the presence of signalized intersections, eco-driving yields significant energy savings by utilizing Signal Phase and Timing (SPaT) information from traffic lights \cite{sun2018robust, bae2018design,Guanetti2018}. 

Our previous work \cite{sun2018robust} focused on the Eco-driving problem through signalized intersections with \emph{uncertain} effective red light duration.
The uncertainty is addressed by formulating chance constraints on passing through the intersections during green lights. 
Simulations showed potential fuel savings of up to 40\%, compared to a modified intelligent driver model \cite{kesting2010enhanced}. That said, the proposed algorithm did not consider surrounding traffic in non-free flow conditions. Also, like most literature on safe and ecological driving, the algorithm was not validated through experiments in real-word traffic conditions. Therefore, in our recent paper \cite{bae2018design}, we extended our previous work to incorporate ACC into the Eco-driving controller. This work balances energy efficiency with collision avoidance and traffic signal compliance. The combined Eco-driving and ACC controller is called Ecological Adaptive Cruise Controller (ECO-ACC). Vehicle-in-the-loop experiments were performed based on a recently introduced test setup for Connected and Automated Vehicles (CAVs) in real-world traffic \cite{kim2018avec,bae2018design}. 

While there is an extensive literature on ECO-ACC, the integration of (\textit{short-sighted}, but aware of immediate traffic changes) ACC and (often \textit{long-sighted}, but only aware of slowly changing traffic information) Eco-driving mostly ends up being a \textit{short-sighted} energy efficient ACC. Therefore, we proposed a \textit{long-sighted} energy efficient ACC in a two-level control framework \cite{bae2018design} that enables both long-term velocity planning and short-term collision avoidance. However, we assumed persistent connectivity with all traffic signals along the route. This allows the ECO-ACC to receive deterministic SPaT information, and an optimal policy can be computed offline. This is hard to implement in practice, due to limited communication range between traffic lights and vehicles. Therefore, one requires an \textit{online} algorithm to recursively update the \textit{long-sighted} velocity trajectory as SPaT information becomes available from approaching intersections.



Since our goal is energy consumption reduction, we are keenly interested in the powertrain dynamics in addition to longitudinal vehicle dynamics control.
In that regard, most existing literature on Eco-driving incorporates a simple powertrain model into its velocity planning \cite{wissam2012optimal, Ozatay2014, sun2018robust}. These existing methods focus on either electric vehicles or gasoline vehicles.
Plug-in hybrid vehicles present an additional challenge, since there are two power generating sources which introduces an additional degree of freedom.
In particular, it is challenging to incorporate a PHEV powertrain model into the Eco-driving optimization problem sinces it increases the state space and control input space size \cite{sciarretta2007control}. To alleviate this issue, an appropriate method needs to be designed to incorporate PHEV powertrain dynamics into the Eco-driving optimization problem, while still retaining sufficient computational simplicity to enable online computations.

The main contributions of this paper address two practical problems. (i) We propose a two-level (velocity planning and safety control) receding horizon control framework that systematically balances energy consumption, travel time, and safety, given limited traffic signal information. (ii) We incorporate a PHEV powertrain model into the velocity planning layer that accurately captures fuel and electricity use for a given powertrain controller, without increasing the state or control input space. This ensures the optimal control problem can be solved in \textit{realtime}. 



The remainder of this paper is organized in the following manner.
Section \ref{sec:controller_design} details the two-level control framework and mathematical formulations of ECO-ACC.
Section \ref{sec:hils} presents simulation results and discusses limitations of the proposed framework and future works.
Section \ref{sec:conclusion} concludes the paper with a summary. 

\section{Receding Horizon Eco-driving Controller Design}
\label{sec:controller_design}
\begin{figure}
    \centering
    \includegraphics[width=1\columnwidth]{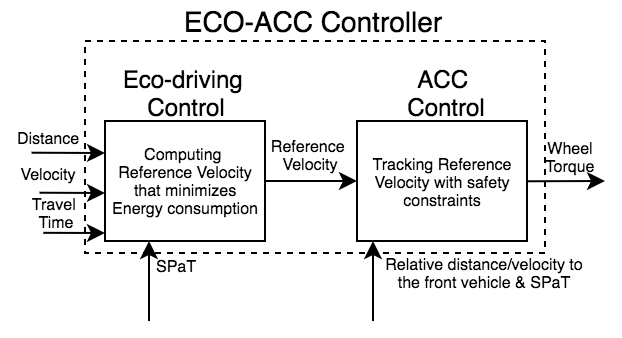}
    \caption{The conceptual diagram of ECO-ACC controller which is composed of the Eco-driving controller and the ACC Controller in separate layers}
    \label{fig:diag_incorporation_controllers}
\end{figure}

We proposed a two-level control framework of ECO-ACC in our previous paper \cite{bae2018design}.
The control architecture is depicted in Fig. \ref{fig:diag_incorporation_controllers}.
The Eco-driving control, which computes a reference velocity trajectory across a long horizon in space, seeks to minimize energy based on both the probabilistic real-time SPaT and its empirical statistics.  
The ACC, which computes the wheel torque to follow the reference velocity, guarantees safety (i.e., collision avoidance and traffic signal compliance) against uncertain road traffic.

In this work we focus on the development of a new Eco-driving control approach, while using the same ACC approach from our previous work \cite{bae2018design}. Readers are encouraged to read \cite{bae2018design} for details about our ACC design.
The novelty of the new proposed Eco-driving control is two-fold: (i) It is executed in a receding horizon control framework with approximated terminal cost and real-time traffic information. (ii) A PHEV powertrain model is incorporated without additional computational costs. 

\subsection{Vehicle Dynamics and Powertrain Model}
In our previous paper \cite{bae2018design}, we only considered longitudinal vehicle dynamics when planning the velocity profile. Consequently, the velocity trajectory is optimized to minimize wheel energy. Although wheel energy is a proxy of fuel and battery energy at a vehicle dynamics level, it does not capture the powertrain dynamics, nor inefficiencies. Therefore, in this paper, we include a powertrain model in both the planning algorithm and HIL simulations to more accurately predict and minimize fuel and battery energy. In this section, we describe the powertrain dynamics as well as vehicle dynamics models used in this work.

\subsubsection{Vehicle Dynamics}
Consider the longitudinal vehicle dynamics where the longitudinal acceleration at step $k$, $a(k)$ is expressed as via Newton's second law of motion
\begin{equation}
    a(k) = \frac{T_w(k)}{mR_{w}}-g\left(cos(\theta(k))C_r -sin(\theta(k))\right) -\frac{\rho AC_{d}}{2m}v(k)^2,\label{eq:acceleration}
\end{equation}
where the input is a wheel torque $T_w(k)$ and the model parameters are vehicle mass $m$, wheel radius $R_{w}$, gravitational acceleration $g$, road grade $\theta$, air density $\rho$, the front cross-sectional area $A$, rolling resistance coefficient $C_r$, and air drag coefficient $C_d$.
The system dynamics are comprised of velocity $v(k)$ and travel time $t(k)$ as states at position $k \Delta s$\footnote{Position $k\Delta s$ is equivalent to step $k$.}, which evolve according to
\begin{equation}
    \underbrace{\begin{bmatrix}
    v(k+1)\\
    t(k+1)
    \end{bmatrix}}_{x({k+1})}
    = \underbrace{\begin{bmatrix}
    v(k)\\
    t(k)
    \end{bmatrix}}_{x(k)}
    + \begin{bmatrix}
    \frac{a(k) \Delta s}{v(k)}\\
    \frac{\Delta s}{v(k) + \frac{a(k) \Delta s}{v(k)}}
    \end{bmatrix}
    \label{eq:dynamics}
\end{equation}
for $k\in\{0,\cdots,N-1\}$, with the position step size $\Delta s$\footnote{Throughout the paper, the position step size is 1 meter, i.e., $\Delta s = 1$.}. We denote $v(k+1) = f_v(k)$ and $t(k+1) = f_t(k)$ for convenience. 

\begin{figure}
    \centering
    \includegraphics[width=1\columnwidth]{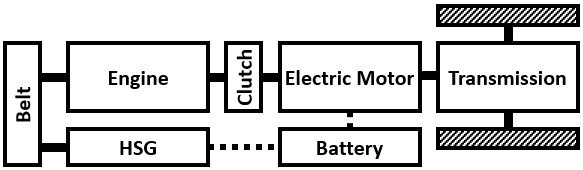}
    \caption{PHEV powertrain architecture}
    \label{fig:PT_model}
\end{figure}

\subsubsection{Powertrain Model}
The powertrain architecture is a pre-transmission parallel hybrid as shown in Fig.~\ref{fig:PT_model}. The input wheel torque $T_{w}(k)$ in the longitudinal vehicle dynamics can be expressed as
\begin{equation}
    T_{w}(k) = r_{gb}(k)T_{\text{sft}}(k)-T_{\text{brk}}(k),
\end{equation}
    where $r_{gb}$ is a transmission gear ratio, $T_{\text{sft}}$ is a shaft torque before the transmission, and $T_{brk}$ is a mechanical braking torque. The shaft torque $T_{\text{sft}}$ is expressed as
\begin{equation}
    T_{\text{sft}}(k) = T_m(k) + e_\text{{on}}(k)\eta_{c}T_e(k),
\end{equation}
where $T_m(k)$ denotes a torque produced by an electric motor at step $k$, $T_e(k)$ is a torque produced by an internal combustion engine at step $k$, $e_{on} \in \{0,1\}$ is the engine on/off status, and $\eta_{c}$ is a clutch efficiency. 

The electric motor power $P_{m}$ can be represented as
\begin{equation}
    P_{m} = \frac{T_m \omega_m}{\eta_m(T_m,\omega_m)}
\end{equation}
where $\omega_m$ is the electric motor speed and $\eta_m$ is an electric motor efficiency, which is a nonlinear function of motor torque $T_m$ and electric motor speed $\omega_m$. The electric motor speed can be computed by $\omega_m = \frac{v}{R_{w}r_{gb}}$. The hybrid starter generator (HSG) power $P_{\text{HSG}}$ and fuel power $P_{f}$ are also computed in the same way as the electric motor power, i.e.,
\begin{equation}
    P_{\text{HSG}} = \frac{T_{\text{HSG}} \omega_{\text{HSG}}}{\eta_{\text{HSG}}(T_{\text{HSG}},\omega_{\text{HSG}})},\quad P_{f} = \frac{T_f \omega_f}{\eta_f(T_f,\omega_f)}.
\end{equation}
Finally, the battery state-of-charge ($SOC$) dynamics can be expressed as 
\begin{equation}
    \dot{SOC} = - \frac{V_{\text{oc}}-\sqrt{V_{\text{oc}}^2-4R_b P_b}}{2R_b Q_b}
\end{equation}
where $P_{b}$ is a terminal battery power, which is written
\begin{equation}
    P_{b} = P_m + P_{\text{HSG}} + P_{\text{aux}},
\end{equation}
$P_{\text{aux}}$ is an auxiliary power, $V_{\text{oc}}$ is an open-circuit voltage, $R_b$ is an internal resistance, and $Q_b$ is a battery pack capacity. Details on PHEV powertrain models can be found in \cite{guzzella2007vehicle}.

\subsection{Cost function for Optimization}\label{sec:cost_function}
Similar to \cite{bae2018design}, the objective is to minimize a convex combination of energy consumption and travel time. However, the differences are the following. First, the control horizon is \emph{limited and receding} and optimal solutions are found \emph{in realtime}. Second, the energy consumption is evaluated based on the powertrain model. Third, SPaT is \textit{uncertain}, except for the upcoming traffic light. 
\subsubsection{PHEV Powertrain Cost Function}
We use a cost function that is the total power cost from the battery and the liquid fuel, mathematically written as
\begin{equation}
\begin{aligned}
    & g_{c}(v(k),T_{w}(k),T_{m}(k),SOC(k)) = \\ & P_{f}(v(k), T_{w}(k) - T_{m}(k)) + s\cdot P_\text{elec}(v(k), T_{m}(k), SOC(k))
\end{aligned}
\label{eq:cost_metric}
\end{equation}
where $s$ is a tuning parameter, and $P_\text{elec}$ is an electrochemical battery power which is computed as $P_\text{elec} = V_{\text{oc}}\cdot I_{b}$ where $I_{b}$ is a current of a battery. Note that the tuning parameter balances the electric power cost with the fuel power cost.

In our simulation studies (details in Section \ref{sec:hils_setup}), we assume that the classic ECMS controller \cite{stockar2011energy, musardo2005ecms} represents the production powertrain controller. Therefore, in our Eco-driving controller, we estimate an output power cost of the ECMS controller as well. We further assume that the ECMS parameter is given and fixed at each SOC level, i.e., the tuning parameter $s$ in \eqref{eq:cost_metric} is deterministic. 


Note that to evaluate the powertrain cost, we additionally need the battery SOC and motor torque, as well as the vehicle dynamics \eqref{eq:dynamics}. These added states mean we must face Bellman's curse-of-dimensionality problem, since we solve the Eco-driving problem \textit{online} via Dynamic Progamming (DP). To remedy the curse of dimensionality, two approximations are made. First, we treat SOC as a fixed parameter over the receding horizon, instead of a dynamic state. Our reasoning is the following: (i) the SOC does not change significantly over a short distance, e.g., few hundreds meters, and (ii) the optimization finds new solutions every few seconds, during which the SOC is reset to its measured value. Second, we approximate powertrain dynamics by a static relation and pre-optimize cost function \eqref{eq:cost_metric} for all possible SOC grid point values with grid step size of 0.01, using the tuning parameter, $s$. That is, we minimize $g_{c}$ from all possible $T_{w}$ and $T_{m}$ combinations. Consequently, the empirical model $g_{c}^\ast$ maps $(v, T_{w}, SOC)$ to a single numeric cost value. 


\begin{figure}
    \centering
    \includegraphics[width=1\columnwidth]{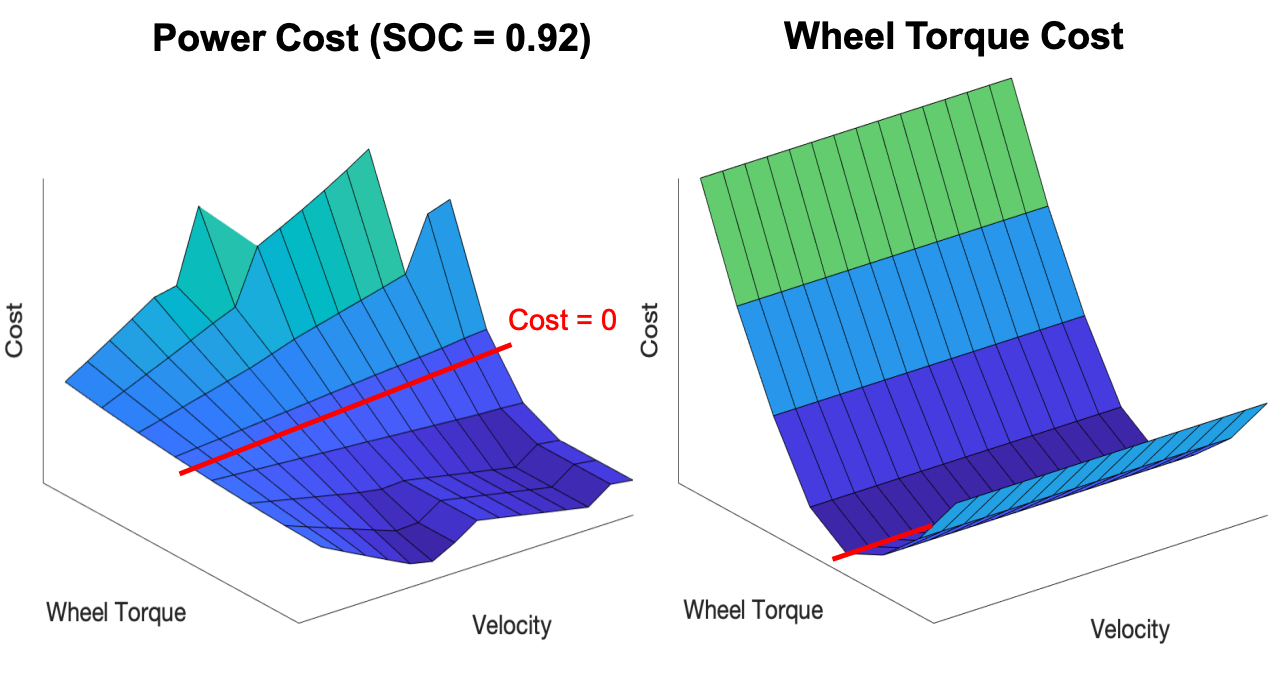}
    \caption{Power map surface. The vertical and horizontal scales of both plots have been omitted for confidentiality reasons. The red line indicates zero cost in each cost function.}
    \label{fig:power_map}
\end{figure}

Fig.~\ref{fig:power_map} shows an example of the obtained empirical cost map (on the left) for the SOC level, 0.92. There are two highlights in the power cost map. (i) Unlike the wheel torque cost map, the power cost map can be negative because of regenerative braking. (ii) Certain wheel torque and velocity combinations result in particularly large power costs. This makes sense because turning the engine on is expensive in terms of energy when the SOC level is high (i.e., when the PHEV is in charge-depleting mode).

To summarize, at the current location $d$ (distance from the origin), the objective function over a receding horizon $d_{H}$ is expressed using the empirical powertrain cost map $g^\ast$ as follows
\begin{equation}
    J_{[d, d_H]} = \sum_{k = d}^{d+d_H} \underbrace{\left(g_{c}^\ast(v(k),T_w(k); SOC(d)) + \lambda \left(\frac{\Delta s}{v(k)}\right)\right)}_{=h(k)}, \label{eq:objective_fun}
\end{equation}
with the weight $\lambda$. The instantaneous cost at each step $k$ is denoted by $h(k)$ for convenience. Note that the second term in $h(k)$ is travel time. Remind that the control variable is the wheel torque $T_{w}(k)$ and we do not consider the allocation problem of engine and motor torque \cite{stockar2011energy, musardo2005ecms}.  

\subsubsection{Approximate Terminal Cost}
The objective function of the cost-minimizing problem from the current location $d$ to destination $d_f$ can be written 
\begin{equation}
    J_{[d,d_f]} = \sum_{i=d}^{d+d_H} h(i) + \sum_{j=d+d_H +1}^{d_f} h(j),\label{eqn:objective_entire_horizon}
\end{equation}
where $h(i)$ denotes an instantaneous cost function at step $i$ in \eqref{eq:objective_fun}. Note that if the information of traffic signal schedules are deterministic and known from the current location to the destination, the controller finds globally optimal solutions. However, the receding horizon controller can utilize the SPaT information within the receding horizon, i.e., only the first term in \eqref{eqn:objective_entire_horizon} can be evaluated. The receding horizon controller is myopic without a terminal cost that captures cost from the end of the receding horizon to the destination. In fact, in our problem where the traffic signal phase and timing is uncertain and dynamic, there is no ``reference" information that represents traffic schedule and flow scenarios. We therefore approximate the expected cost beyond the current receding horizon as a sample mean over different scenarios. 
The objective function is re-written 
\begin{equation}
    \min \sum_{i=d}^{d+d_H} h(i) + \hat{J}_{[d+d_H+1, d_f]},\label{eq:approx_terminal_cost}
\end{equation}
where 
\begin{equation}
\hat{J}_{[d+d_H+1, d_f]} = \sum_{j=d+d_H +1}^{d_f} \mathbb{E}[h(j)].
\end{equation}
The expectation is taken w.r.t. uncertain SPaT. Using a sample mean over randomly generated scenarios, the expected terminal cost reads
\begin{equation}
    \hat{J}_{[d+d_H+1,d_f]} = \frac{1}{M} \sum_{j=1}^M \min \sum_{i=d+d_H+1}^{d_f} h(i; \sigma_j),
\end{equation}
where $M$ is the total number of randomly generated SPaT scenarios and $\sigma_j$ represents SPaT scenario $j$. This approximation enables us to construct a terminal cost function that accounts for cost-to-go beyond the control horizon, while accounting for uncertain SPaT.


\subsubsection{Soft terminal constraint}
In addition to the approximated terminal cost \eqref{eq:approx_terminal_cost}, we leverage soft constraints to penalize deviations from a reference travel time. The soft terminal constraint in each receding horizon $d_H$ is written
\begin{equation}
    \underbrace{\big[t^D_f - (t+\tau_H)\big]}_{\text{Remaining time}}\hat{v}_{\text{avg}} \geq \underbrace{\big[d_f-(d+d_H)\big]}_{\text{Remaining distance}}-\gamma \label{eq:soft_const}
\end{equation}
where $\tau_H$ is a total travel time within the receding horizon, $t_f^D$ is a desired travel time over the receding horizon, $\hat{v}_{\text{avg}}$ is an empirical average speed over the remaining route, and $\gamma$ is a slack variable. Note that we use the slack variable $\gamma$ in \eqref{eq:soft_const} to make the constraint ``soft'' since each traffic scenario is randomly generated and therefore a hard constraint may result in infeasible solutions. The slack variable is evaluated in the terminal cost as
\begin{equation}
    \gamma = \big[d_f-(d+d_H)\big]-\big[t^D_f - (t+\tau_H)\big]\hat{v}_{\text{avg}}.
\end{equation}
Without this substitution, we require an additional control variable for the slack variable, which correspondingly increases the problem dimensions. 

\subsection{Formulation of Constraints}
The constraints are set to
\begin{align}
    T_{w}^{\text{min}} \leq T_{w}(k) &\leq T_{w}^{\text{max}}, \label{eq:ineq_whl_trq}\\
    a^{\text{min}}\leq a(k)&\leq a^{\text{max}}, \label{eq:ineq_a}\\
    v^{\text{min}}(k)\leq v(k) &\leq v^{\text{max}}(k), \label{eq:ineq_v}\\
    t^{\text{min}}(k)\leq t(k) &\leq t^{\text{max}}(k), \label{eq:ineq_t}
\end{align}
for all $k\in\{d,\cdots, d+d_{H}\}$, where the inequality constraints \eqref{eq:ineq_whl_trq}, \eqref{eq:ineq_a}, and \eqref{eq:ineq_v} ensure that the wheel torque, acceleration, and velocity, respectively, are bounded. Particularly, the wheel torque $T_{w}$ is lower-bounded by the maximum braking torque which is the sum of the maximum mechanical friction and regenerative braking torques. Both the maximum wheel torque and regenerative braking torque are governed by characteristics of the electric motor, and their values depend upon the shaft speed, i.e., the vehicle speed. The maximum acceleration $a^{\text{max}}$ is set to a physically feasible limit, and the maximum velocity $v^\text{max}$ is set to the maximum speed limit on the road. The inequality constraint \eqref{eq:ineq_t} ensures that the travel time is bounded by minimum and maximum travel time boundaries ($t^\text{min}(k)$ and $t^\text{max}(k)$, respectively).

\subsubsection{Dynamic constraints for traffic lights}\label{sec:dynamic_constraints}
We utilize the SPaT information in the form of constraints. We assume that the current SPaT of the next traffic light is given, however, only historical SPaT of the other traffic lights are given. This assumption corresponds to the limited range of V2I communication in practice. 

Given SPaT information of the next traffic light, we find ``infeasible" cases. The main intuition of the infeasible cases at each intersection is that the vehicle cannot pass through (or, ``infeasible" to pass through) the intersection during the red light phase (i) in the current cycle; and (ii) in the next cycles. 
When the current signal phase is yellow, the controller is set to be conservative so that the car does not pass through the intersection, for safety reasons. Mathematically, we find the ``infeasible" set (denoted by $\mathbf{IS}$) of states $x(k) = [v(k),t(k)]^\top$, for step $k\in\{d,...,d+d_H\}$, that satisfy logical conditions:

{\small
\begin{equation*}
\begin{cases}
(f_t(k)\leq s_t) \cup \bigg\{\left(f_t(k)>s_t\right) \\ \cap \left(\mathbf{R}\left(f_t(k)-s_t,\ell^{(n)}_c\right)\geq \ell^{(n)}_c-\hat{\ell}^{(n)}_r \right)\bigg\} & \text{if }s_p = \text{red}\\
\hline
(f_t(k)>s_t) \\\cap (\mathbf{R}(f_t(k)-s_t,\ell^{(n)}_c)\leq \hat{\ell}^{(n)}_r) & \text{if } s_p = \text{green}\\
\hline
(f_t(k)\leq s_t)\cup \bigg\{(f_t(k)>s_t) \\\cap (\mathbf{R}(f_t(k)-s_t,\ell^{(n)}_c)\leq \hat{\ell}^{(n)}_r)\bigg\} & \text{if } s_p = \text{yellow}
\end{cases}    
\end{equation*}
} if an intersection is located at step $k+1$ and it is the first upcoming intersection from the current location $d$, and
\begin{equation}
\mathbf{R}(f_t(k)+\ell^{(n)}_{c,O},\ell^{(n)}_c)\leq \hat{\ell}^{(n)}_r, \label{eq:infeas_modulor}
\end{equation}
if an intersection is located at step $k+1$ and it is not the first upcoming intersection, where $s_p$ is a current signal phase, $s_t$ is a remaining time of the signal phase, $\ell_c^{(n)}$ is a signal cycle length at intersection $n$, $\hat{\ell}^{(n)}_r$ is an estimated red light duration, $\ell_{c,O}^{(n)}$ is a time shift of the signal cycle initiation, and $\mathbf{R}(\cdot)$ is the modulo operator. Recall that $f_t(k)$ is a travel time at step $k+1$ given the states and input pair $(x(k),T_w(k))$ at step $k$.

\begin{figure}
    \centering
    \includegraphics[width=1\columnwidth]{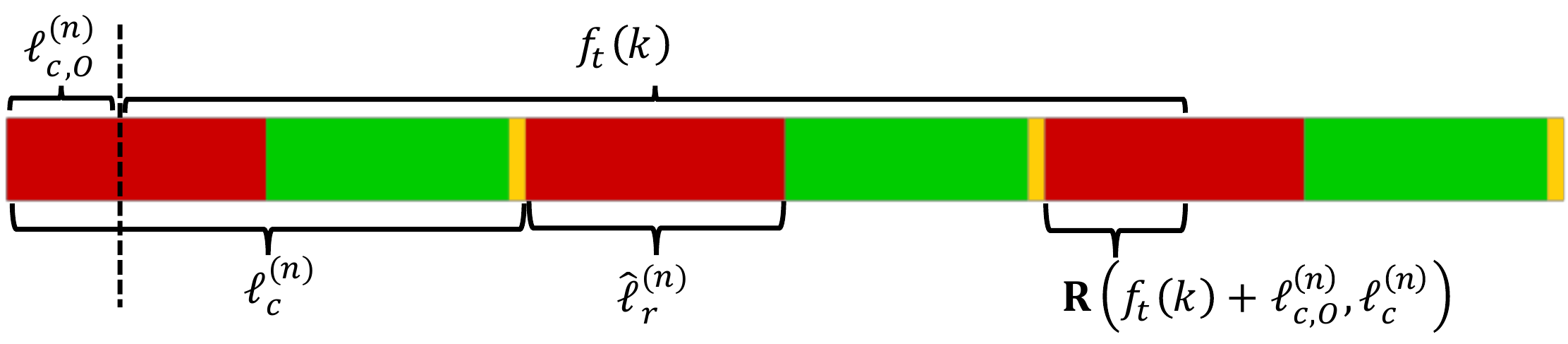}
    \caption{Graphical demonstration of a cycle clock time $\mathbf{R}(f_t(k)+\ell^{(n)}_{c,O},\ell^{(n)}_c)$. Each colored block represents a traffic light, i.e., red, green, and yellow from left.}
    \label{fig:infeas_set}
\end{figure}

The inequalities with the modulo operator $\mathbf{R}(\cdot)$ indicate the following. At each intersection, given current states (velocity and travel time at step $k$), it is infeasible if a travel time at step $k+1$ is within a red light period of a traffic signal cycle. Take the inequality \eqref{eq:infeas_modulor} as an example. The left hand side represents the remainder of ``the sum of cumulative travel time and shifted cycle initiation time" divided by ``the signal cycle length". The remainder is equivalent to a clock time within the signal cycle, as depicted in Fig.~\ref{fig:infeas_set}. If the cycle clock time is less than the estimated red light duration, i.e., less than $\hat{\ell}^{(n)}_r$, then the state and input pair $(x(k),T_w(k))$ is set to infeasible. Similarly, we find infeasible states at the upcoming intersections with deterministic SPaT.

The estimated red light duration $\hat{\ell}^{(n)}_r$ is determined as a score at $\eta^{th}$ percentile of the conditional probability density function (PDF) of red light durations. The optimal wheel torques at the infeasible states are forced to be minimum, i.e., maximum braking, to ensure the vehicle does not pass the intersection during a red light.

\subsection{Receding Horizon Control Formulation}
The complete optimization problem is summarized as
\begin{equation}
    \min_{T_w, v, t} \quad J_{[d,d_H]} + \hat{J}_{[d+d_H+1,d_f]} + \beta\gamma^2\label{eq:opt_obj}
\end{equation}
subject to 
\begin{align}
    &\text{vehicle dynamics \eqref{eq:dynamics}}\nonumber\\
    &\text{constraints \eqref{eq:ineq_whl_trq}-\eqref{eq:ineq_t}}\nonumber \\
    &\text{feasible states } x; x \notin \mathbf{IS}. \nonumber
\end{align}
We apply dynamic programming (DP) to solve the above optimization problem, given the current states and SPaT information from the next traffic light. Algorithm \ref{alg:opt_wheel_torque} summarizes the process of computing the optimal wheel torques $T_w^\ast$ at each iteration (remind that $T_w^\ast$ represents a map of optimal wheel torque associated with states, i.e., $T_w^\ast$ denotes a policy map). 
The complete algorithm is the following. First, the Eco-driving controller receives the current information (states and SPaT). Second, the above optimization problem is solved via Algorithm \ref{alg:opt_wheel_torque}, which takes few seconds, and the policy map $T_w^\ast$ is updated with the recent policy map. It is important to note that while the optimization is being solved with the recently measured states and SPaT, the Eco-driving controller sends the ACC controller a velocity reference associated with the current states, i.e., $T_w^\ast(d,v,t)$, at every $0.2$ seconds. The Eco-driving controller repeats solving the optimization problem until the vehicle arrives at the destination, i.e., $d\leq d_f$.

In Algorithm \ref{alg:opt_wheel_torque}, $d_I$ is a set of traffic light locations, ($n_v\times n_t$) is the grid size of ($v$,$t$), the operator $(\cdot)_+$ takes a positive element in $(\cdot)$, and $\beta$ is the slack variable weight $\gamma$ defined in Section \ref{sec:cost_function}.


\begin{algorithm}
    \SetKwInOut{Input}{Input}
    \SetKwInOut{Output}{Output}
    \SetKwInOut{Init}{Init}
    
    \Input{$d,t,d_{I},d_{H},\ell_c,\hat{\ell}_r,s_p,s_t$}
    \Output{$T_w^\ast \in \mathrm{R}^{(d_H\times n_v\times n_t)}$}
    \Init{Compute relative distance to traffic lights within the distance horizon\\$\tilde{d}_{I}=(d_{I}-d)_+ \in [0, d_H]$\\
    
    \vspace{3mm}
    Set the terminal cost\\
    $\hat{J}_{[d+d_H+1,d_f]} + \beta \gamma^2$}
    \For{$k=d_{H}-1 \rightarrow 0$}{
    Solve Bellman's equation for all feasible states
    $\forall(v_i,t_j) \in \{\left(v(k),t(k)\right) | \left(v(k),t(k)\right) \notin \mathbf{IS} \},$
    
    $V_k(v_i,t_j) = \min_{T_w}\{g(v_i,t_j)+V_{k+1}(f(v_i,t_j,T_w))\}$
    
    \vspace{3mm}
    Get the minimizers
    $[T_w^\ast]_{i,j} \leftarrow \text{minimizer of } V_k(v_i,t_j)$
    }
    \caption{Computing optimal wheel torques}\label{alg:opt_wheel_torque}
\end{algorithm}

\section{Hardware-in-the-loop Simulation Result}\label{sec:hils}
\subsection{Hardware-in-the-loop Simulation setup}\label{sec:hils_setup}
The hardware-in-the-loop simulation (HILS) is identical to our previous work \cite{bae2018design}, except that the real PHEV is replaced by mathematical models. 
Therefore, the HILS in this paper consists of a desktop (for traffic simulation and for the mathematical vehicle model), Matrix embedded PC-Adlink (for Eco-driving control), and dSpace MicroAutoBox (for ACC). Note that ACC updates its torque control every 0.2 seconds to ensure safety from immediate traffic changes. The Eco-driving control updates its solution every ~4 seconds to find a local optimal velocity profile over a few hundred meters\footnote{One can adjust the receding control horizon for the Eco-driving controller to compromise the computation time with solution optimality.} ahead. While the Eco-driving controller is computing a new solution with recent traffic information, the optimal velocity at current distance and travel time is found from the most recent solution and sent to the ACC at every recalculation of optimal torque control, i.e., 0.2 seconds. As in \cite{bae2018design}, we consider the Live Oak corridor in Arcadia, California, with a total of eight intersections, for our simulation study.



\subsection{Simulation Result}
\begin{figure}
    \centering
    \includegraphics[width=1\columnwidth]{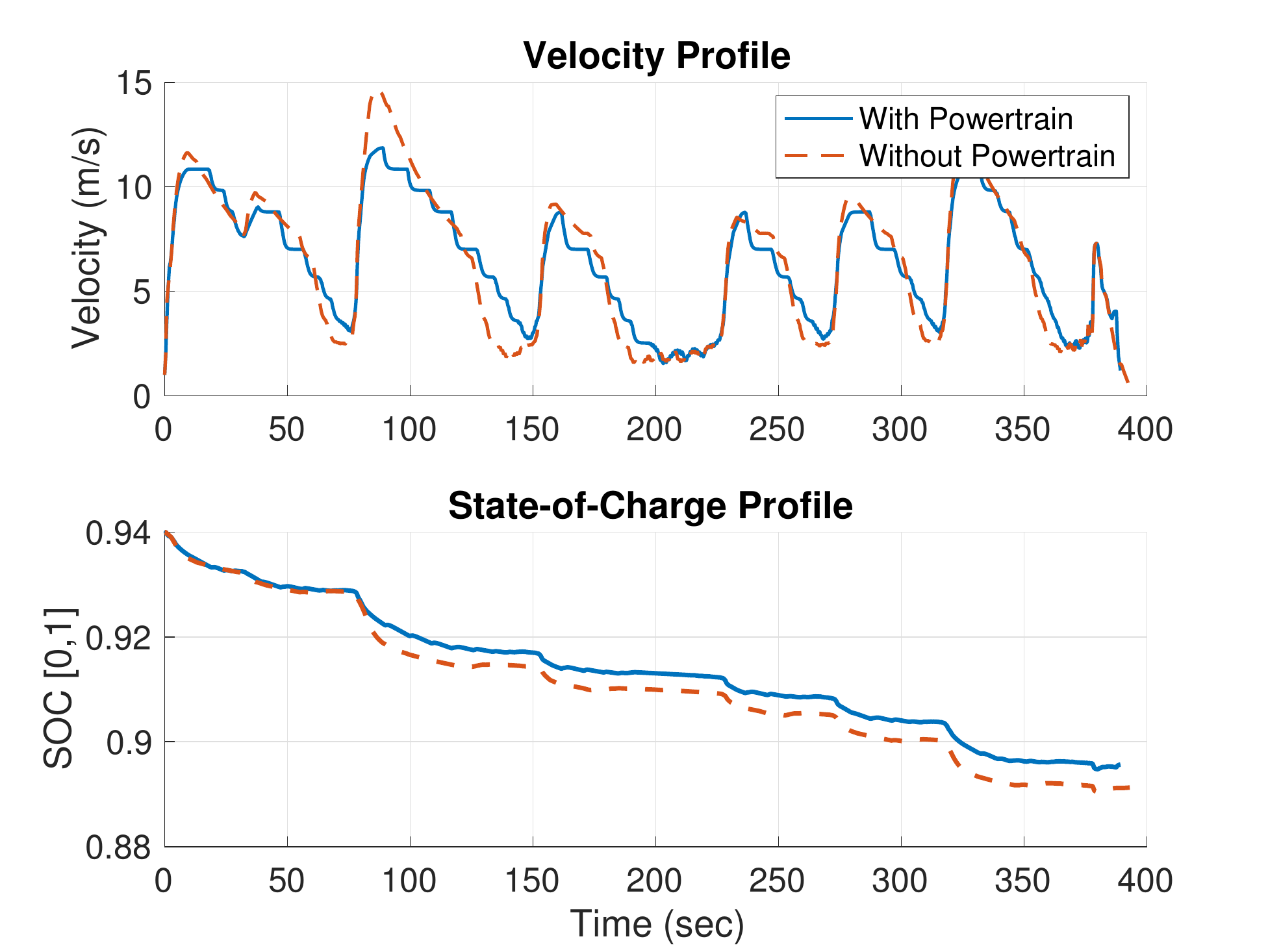}
    \caption{Comparisons in Velocity and State-of-Charge of ECO-ACC with powertrain and without powertrain.}
    \label{fig:soc_pt_vs_without}
\end{figure}

In this section, we present simulation results to illustrate the effectiveness of the proposed control framework. 
We first validate the power cost map used in the Eco-driving controller in terms of energy savings, compared to the wheel torque map (Fig.~\ref{fig:power_map}). 
We consider a deterministic SPaT scenario to keenly evaluate (ideal) energy savings. Fig.~\ref{fig:soc_pt_vs_without} illustrates velocity and SOC trajectories for the two controllers. Only SOC profiles are shown because the engine remained off in both cases. The controller using the power cost map benefits from occasional instantaneous negative power, meaning that it strategically uses regenerative braking to conserve battery SOC. 
As a result, the controller with the power cost map spends less battery energy, improving energy performance by 9.02\% in MPGe.

\begin{figure}
    \centering
    \includegraphics[width=1\columnwidth]{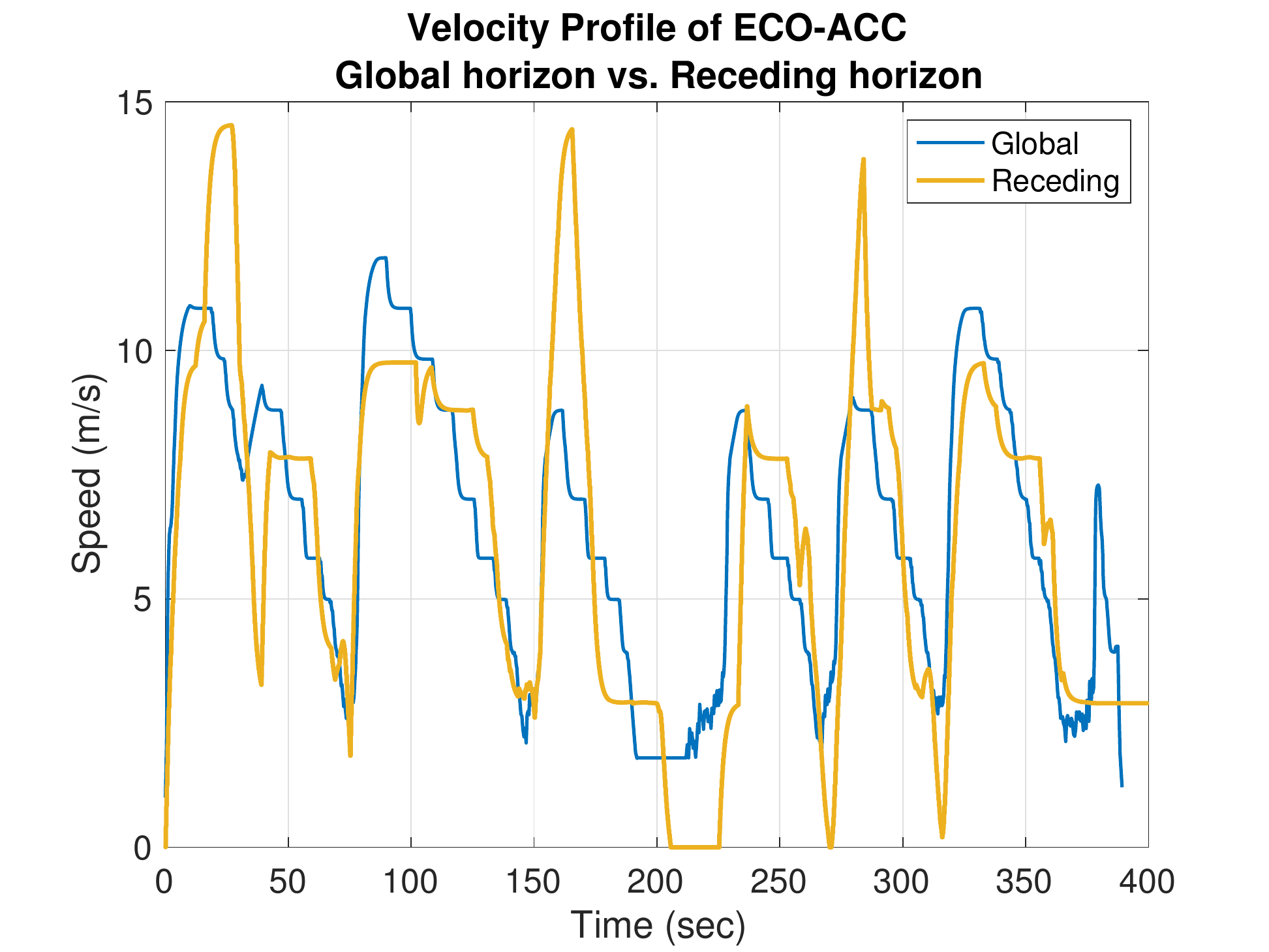}
    \caption{Comparison in actual velocity profile }
    \label{fig:receding_vs_global}
\end{figure}

We then examine the energy consumption of the receding horizon ECO-ACC compared to that of the global horizon ECO-ACC. Note that the global horizon ECO-ACC has perfect information of SPaT over all intersections, and therefore DP finds a global optimal velocity profile from origin to destination, $2500$ meters. In contrast, the receding horizon ECO-ACC finds a locally optimal velocity profile over the receding horizon, which is set to a V2I communication range limit, $400$ meters. 
Fig.~\ref{fig:receding_vs_global} shows the velocity profiles of ECO-ACC. The velocity profile of the global horizon ECO-ACC illustrates that, in the ideal case, the vehicle does not have to speed up to its maximum and it does not have to stop at the intersections in the middle of the route. In contrast, the velocity profile of the receding horizon ECO-ACC is more volatile, approaching the maximum speed limit and zero speed. This is because the receding horizon controller is myopic and the SPaT information is limited to the next intersection. Consequently, the energy efficiency in MPGe is 14.97\% lower in the receding horizon control compared to the global horizon control. 

\begin{figure}
    \centering
    \includegraphics[width=1\columnwidth]{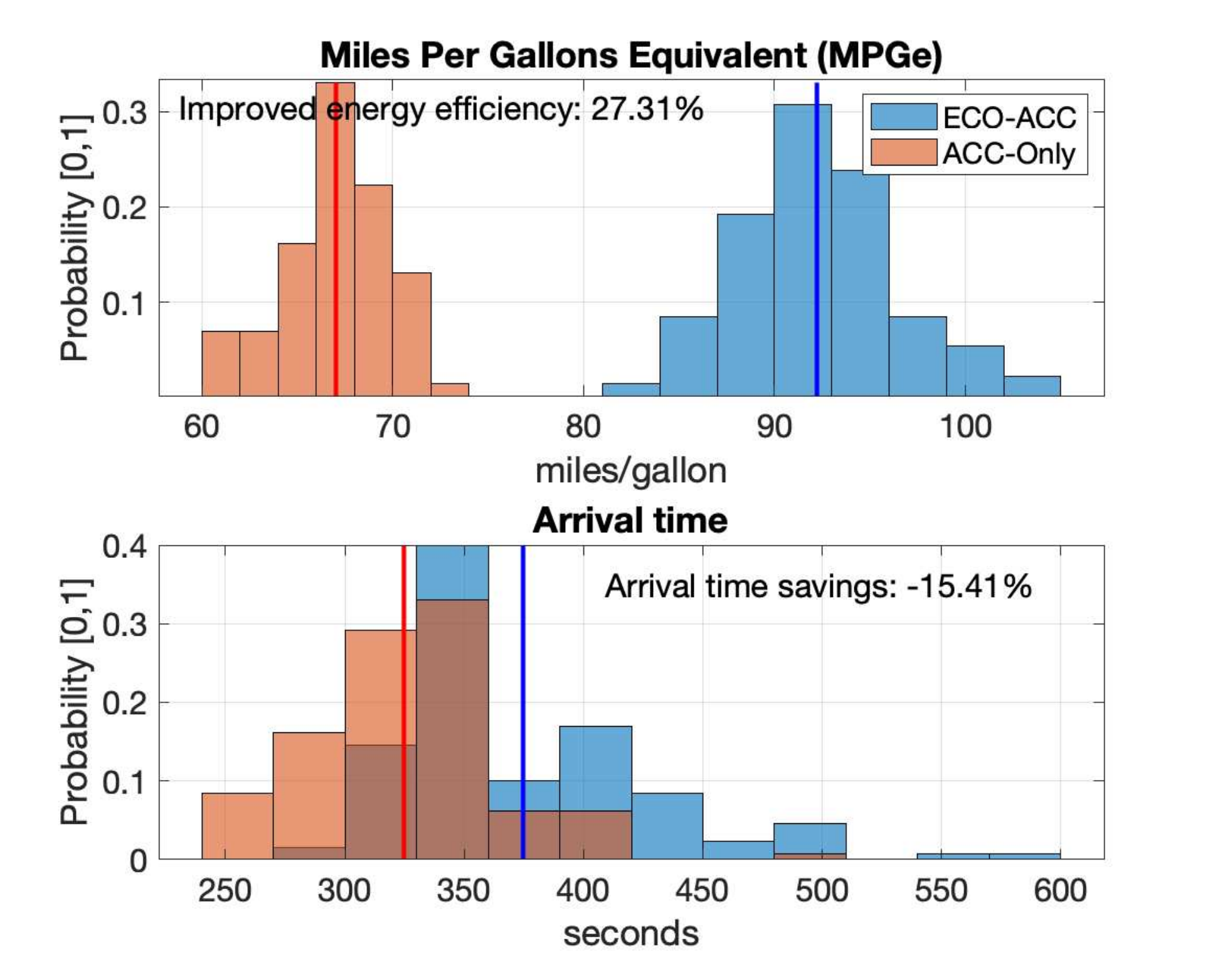}
    \caption{The probability distributions of MPGe (top) and arrival times (bottom) obtained by Hardware-in-the-loop Simulations. The distributions consist of a total of 130 traffic scenarios in each ECO-ACC and ACC-Only cases. We calculate that 33.7 kilowatt-hours of electricity is equivalent to one gallon of gas \cite{friedman2014analysis}.}
    \label{fig:mpge}
\end{figure}

Finally, we validate the energy savings of the receding horizon ECO-ACC in various traffic scenarios, based on Monte Carlo simulations. At each simulation, traffic schedules (e.g., red light duration, time shift of cycle initiation at each intersection, positions of other vehicles) are randomly sampled from empirical PDFs. To compute the conditional PDF we use a month of SPaT data collected by Sensys Networks over the Live Oak route in Arcadia. In Fig. \ref{fig:mpge}, it is clearly seen that ECO-ACC is significantly more energy efficient ($27.31\%$) compared to ACC only (ACC with constant velocity reference), in any traffic scenarios at any hour of the day. That said, ECO-ACC results in longer travel time ($15.41\%$) compared to ACC only, which is aligned with the intuition from our previous experimental results with the global horizon ECO-ACC in \cite{bae2018design}. One can further investigate with different penalties on travel time in the objective function \eqref{eq:objective_fun} to trade off energy consumption with travel time. 

\subsection{Limitations and Future Work}
One limitation of the receding horizon control framework is that the Eco-driving controller can possibly update a DP solution \textit{immediately before} passing the intersection. Consequently, the velocity planning will only rely on the historical SPaT at the next intersection, which might result in poor performance if the historical SPaT does not represent the actual SPaT. If the estimated SPaT of the next intersection has a large offset with the actual SPaT, then ECO-ACC can unnecessarily waste energy. For example, suppose that the optimal velocity plan is to keep a current velocity until the next intersection. With a poorly estimated SPaT, the vehicle may reduce its velocity until the DP is recalculated with an updated, actual SPaT. Consequently, the vehicle needs to spend additional energy to catch up with the optimal velocity trajectory. These offsets are demonstrated as damping points around 100 or 255 seconds in Fig.~\ref{fig:receding_vs_global}. Therefore, a robust design for velocity trajectories remains as a challenge.

Another limitation is that we only considered a charge-depleting powertrain mode in the simulation (SOC level starts high). Due to this assumption, the liquid fuel usage was negligible, and total energy consumption heavily depends on the battery usage. That being said, it is straightforward to extend the proposed framework to adopt a charge sustaining mode by considering powertrain cost maps parameterized by SOC, which also remains for future work.







\section{Conclusion}\label{sec:conclusion}
This paper proposes a receding horizon control framework for an online Ecological Adaptive Cruise (ECO-ACC) control, with considerations for limited vehicle-to-infrastructure communication range and energy consumption behavior for Plug-in Hybrid Vehicles (PHEVs). The overall objective is to minimize energy consumption while avoiding collisions and complying with traffic signals. The framework is based on a two-layer structure, where the upper layer corresponds to the velocity planning algorithm and the lower layer corresponds to collision avoidance and traffic signal compliance. This paper focuses on the velocity planning algorithm in the upper layer, which is adaptive to dynamically updated traffic signals within a receding control horizon. The receding control scheme is designed for hardware implementation and experimentation. Several practical issues were addressed, including efficient computations and limited traffic signal information \textit{realtime}. Our control design is experimentally validated through a recently developed hardware-in-the-loop simulation.


\section*{Acknowledgement}
The information, data, or work presented herein was funded in part by the Advanced Research Projects Agency-Energy (ARPA-E), U.S. Department of Energy, under Award Number DE-AR0000791. The views and opinions of authors expressed herein do not necessarily state or reflect those of the United States Government or any agency thereof.

The authors would also like to thank Hyundai America Technical Center, Inc. for providing us with the vehicle and a testing facility, and Sensys Networks for granting access to the traffic data used in this paper.


\bibliography{ref}


\vfill

\end{document}